\documentclass[%
 aip,
 amsmath,amssymb,
 reprint,%
]{revtex4-2}

\usepackage{graphicx}
\graphicspath{ {../Figures/} }
\usepackage{dcolumn}
\usepackage{bm}

\usepackage[utf8]{inputenc}
\usepackage[T1]{fontenc}
\usepackage{mathptmx}
\usepackage{etoolbox}

\usepackage[colorlinks=true, allcolors=blue]{hyperref}
\usepackage{url}

\usepackage{siunitx}

\begin{document}

\title{Step-by-step design guide of a cryogenic three-axis vector magnet}\thanks{This work was published in \href{https://doi.org/10.1063/5.0270187}{Rev.\ Sci.\ Instrum.\ \textbf{96}, 065208 (2025).}}

\author{Gaia Da Prato}
\affiliation{Kavli Institute of Nanoscience, Department of Quantum Nanoscience, Delft University of Technology, 2628CJ Delft, The Netherlands}

\author{Yong Yu}
\affiliation{Kavli Institute of Nanoscience, Department of Quantum Nanoscience, Delft University of Technology, 2628CJ Delft, The Netherlands}

\author{Ronald Bode}
\affiliation{Kavli Institute of Nanoscience, Department of Quantum Nanoscience, Delft University of Technology, 2628CJ Delft, The Netherlands}

\author{Simon Gr{\"o}blacher}
\email[]{s.groeblacher@tudelft.nl}
\affiliation{Kavli Institute of Nanoscience, Department of Quantum Nanoscience, Delft University of Technology, 2628CJ Delft, The Netherlands}


\begin{abstract}
A tunable magnetic field at low temperatures is essential for numerous applications, including spintronics, magnetic resonance imaging, and condensed matter physics. While commercial superconducting vector magnets are available, they are complex, expensive, and often not adaptable to specific experimental needs. As a result, simple in-house designs are often being used in research environments. However, no comprehensive step-by-step guide for their construction currently exists. In this work, we provide a detailed manual for designing and building a cryogenically compatible three-axis vector magnet. The system is tested at the mixing chamber of a dilution refrigerator at temperatures ranging from 15~mK to 4~K, with no significant increase in base temperature. Safety measures are implemented to mitigate heating from quenching. The coils are successfully driven with DC currents as high as 3~A, generating magnetic fields of up to 2.5~T in the bobbin's bore and 0.4~T at the sample position. Magnetic field measurements using Hall sensors demonstrate good agreement with the predictions of the designed performance.
\end{abstract}


\maketitle

\section{Introduction}
\label{sec:Introduction}
Tunable magnetic fields at low temperatures are essential for both fundamental research and practical applications. They enable precise control of spin systems, such as rare-earth-doped crystals~\cite{Raha2020, Rochman2023, Strinic2025}, color centers in diamond~\cite{Whaites2023} or in silicon carbide~\cite{Wang2023}, and quantum dots~\cite{Huber2019}. Additionally, low-temperature magnetic fields play a crucial role in tuning microwave resonators~\cite{Xu2019}, investigating quantum phase transitions~\cite{Lee2022}, and studying low-temperature magnonics~\cite{Cocconcelli2024} and other condensed matter phenomena~\cite{Pan2020, Pekola2021}. On the application side, superconducting magnets provide stable high magnetic fields for technologies such as magnetic resonance imaging~\cite{Wang2023_2, Lvovsky2013, Deguchi2004}.

Since permanent magnets offer no tunability without mechanical movement, electromagnets -- such as coils -- are widely used. In many cases, controlling not only the field strength but also its direction is essential~\cite{Raha2020, Chen2024, Rossi2012}. A straightforward approach involves rotating either the magnet or the sample using a motorized stage~\cite{Holzapfel2024, Izawa2001, Suderow1997, Hess1994}. However, such mechanical motion introduces vibrations, which can disrupt sensitive experiments~\cite{Galvis2015, Zhang2024}. Furthermore, movement within a cryogenic environment can generate unwanted heat, making this method unsuitable for ultra-low-temperature setups, particularly in dilution refrigerators operating at tens of Millikelvin. In such cases, a vector magnet - capable of producing a controlled magnetic field in an arbitrary direction without mechanical motion - is indispensable. Deploying such electromagnets inside a cryostat presents significant thermal management challenges however. Resistive wires are impractical due to excessive heat dissipation, necessitating the use of superconducting wires. Superconducting magnets have been extensively studied~\cite{Stekly1965, Wilson1983, Iwasa2009}, but minimizing cryostat heating and preventing magnet quenches remain critical engineering concerns.

Although commercial vector magnets exist~\cite{BlueforsVectorMagnet, CryomagneticsVectorMagnet, OxfordInstrumentsMagnets}, they are typically expensive, very complex, and are often designed for specific cryostat models, limiting flexibility. Consequently, many research groups have developed their own vector magnet solutions~\cite{Raha2021, Rochman2023, Weaver2024, Etesse2021, Strinic2025, Liu2020, Zhang2021, Galvis2015}. However, no comprehensive guide exists for designing and constructing such systems, forcing each research group to invest significant time in the development process. Furthermore, existing solutions often lack one or more key features, including:
\begin{enumerate}
\item Operation across a wide temperature range (from 15~mK to 4~K)
\item Generation of high magnetic fields (up to 2.5~T) with relatively low currents (3~A)
\item Three-axis control for full vector manipulation
\item Measures to minimize heating and prevent quenches
\item Simulation for magnetic field estimation in three dimensions.
\end{enumerate}
In this work, we present the design, construction, and optimization of a superconducting three-axis vector magnet operating at the mixing chamber (MXC) plate of a dilution refrigerator (DR). Our system meets all the above criteria and offers a detailed, step-by-step guide for reproducibility and customization. Specifically, our vector magnet is designed for spectroscopic studies of erbium-doped silicon waveguides. The sample is thermally anchored at the MXC plate of the DR and the light is coupled to the waveguides via a lensed fiber. To be able to address different devices and to optimize the coupling, the fiber is mounted on a three-axis nanopositioner (see Figure~\ref{fig:B_field_vs_parameters}(d)). Our vector magnet design therefore also accounts for the nanopositioner and sample holder, which can be inserted in the cross section of the axes of the three magnets.

The paper is structured as follows:\ Section~\ref{sec:sim} introduces our model for calculating and optimizing the generated magnetic field. Section~\ref{sec:assembling} describes coil construction and thermalization within the DR. Section~\ref{sec:safety} details quench protection mechanisms. Section~\ref{sec:mag_field_meas} compares theoretical predictions with experimental measurements. Finally, Section~\ref{sec:Conclusion} summarizes our findings and discusses potential improvements.

\section{Simulation and design}
\label{sec:sim}

In order to develop our three-axis vector magnet, we first estimate the magnetic field generated by a single coil and then extend the analysis to the full system. The design parameters are then optimized to achieve the highest possible magnetic field at a fixed current.

\subsection{Calculation of the magnetic field}
\begin{figure}[htb!]
\centering
\includegraphics[width = 0.9 \linewidth]{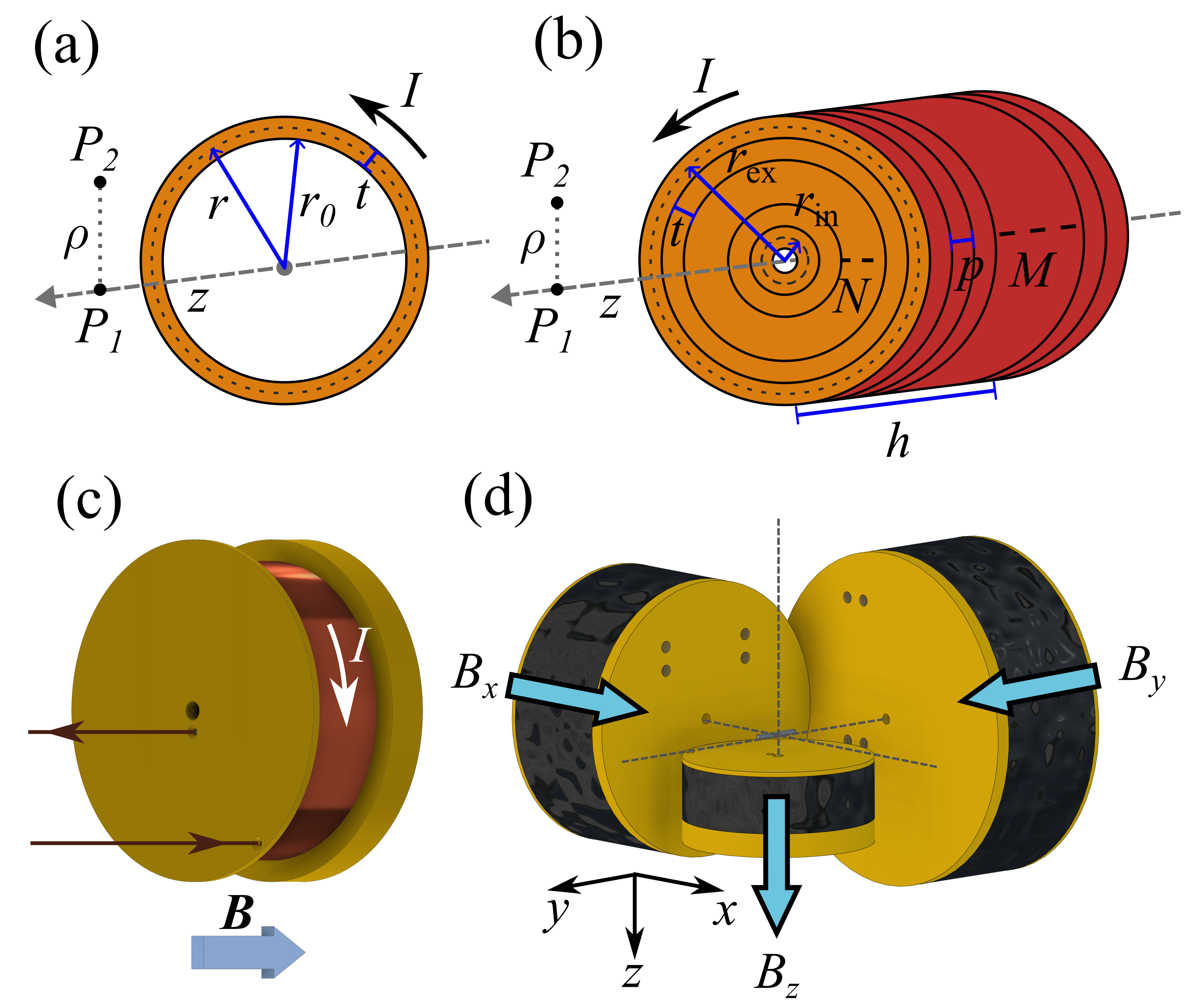}
\caption{(a) Current loop schematic. (b) Coil model used for the calculation. (c) Schematic of a wound bobbin showing the holes where the wire enters and exits the bobbin. (d) Three-axis vector magnet with the coordinate system used throughout this work.}
\label{fig:calculation}
\end{figure}

We employ an analytical approach to estimate the magnetic field generated by a coil. Considering a single loop of radius $r$ carrying a DC current $I$ we then calculate the magnetic field $\textbf{B}$ at a point on the loop axis, at a distance $z$ from its center. Using the law of Biot-Savart, we find the well-known result~\cite{Griffiths2017}:
\begin{equation}
\label{eq:single_loop}
\textbf{B}(\textbf{z}) = \frac{\mu_0 I}{2}\frac{r^2}{(r^2+z^2)^{3/2}} \hat{\textbf{z}},
\end{equation}
where $\mu_0 = 4\pi \times 10^{-7}$ N/A\textsuperscript{2} is the vacuum permeability.\\

In a more realistic model, the loop consists of a wire with finite thickness $t$, internal radius $r_0$ and external radius $r_0+t$, as illustrated in Figure~\ref{fig:calculation}(a). The field can then be approximated using Eq.~\eqref{eq:single_loop} with $r=r_0$ (see Appendix~\ref{app:calc_magnetic}). The coil is characterized by five parameters:\ the internal radius $r_{\mathrm{in}}$, the external radius $r_{\mathrm{ex}}$, the height $h$, the wire thickness $t$, and the pitch $p$, which defines the spacing between adjacent loops (see Figure \ref{fig:calculation}(b)). We model the coil as $N$ concentric loops of thickness $t$, repeated $M$ times adjacently. Here, we are assuming that the layers stack one above another, even though likely they stack in a hexagonal packing arrangement~\cite{Raha2021}. We can do this simplification because the translation between two consecutive layers, given by $t/2$, is negligible and the variation in distance between two stacked layers -- roughly $15\%$ -- is also irrelevant in a real application. 
The total magnetic field at a point $P_1$ on the axis, at a distance $z$ from the closest loop, is then given by:
\begin{equation}
\label{eq:B_bobbin_axis}
B_{\mathrm{coil}}(z, r_{\mathrm{in}}, r_{\mathrm{ex}}, h, t, p, I) = \sum_{i = 1}^{N}\sum_{j = 1}^{M}B_{\mathrm{loop}}(z_j,r_i,I),
\end{equation}
where
\begin{equation}
\left\{\begin{array}{l}
N = \mathrm{floor}[(r_{\mathrm{ex}}-r_{\mathrm{in}})/t]\\
r_i(r_{\mathrm{in}}, t) = r_{\mathrm{in}} + \sum_{k = 0}^{j}t \cdot k\\
M = \mathrm{floor}[h/p]\\
z_j(z, p) = z + \sum_{k = 0}^{j}p \cdot k\\
\end{array}\right.
\end{equation}
and the magnetic field generated by a single loop is
\begin{equation}
B_\mathrm{loop}(z_j,r_i,I) =  \frac{\mu_0 I}{2}\frac{r_i^2}{(r_i^2+z_j^2)^{3/2}}.
\end{equation}

To optimize the coil parameters, knowledge of the on-axis field is sufficient in our case, as the sample is positioned at the intersection of the three axes of the coils. However, for practical considerations, such as estimation of misalignment effects (see Section~\ref{sec:effective_parameters}) and field measurements using off-axis Hall probes (see Section~\ref{sec:mag_field_meas}), it is necessary to compute the field at arbitrary points.

Given a point $P_2$ at distance $z$ from the loop and $\rho$ from the axis (see Figure~\ref{fig:calculation}(a)), the axial and radial components of the magnetic field are respectively given by~\cite{Dennison_MagnetFormulas}:
\begin{equation}
\mathbf{B}_{\mathrm{loop},z}(z,\rho, r, I) = B_0 \frac{1}{\pi Q}\left[E(k)\frac{1-\alpha^2-\beta^2}{Q-4\alpha}+K(k)\right]\mathbf{\hat{z}}
\end{equation}

\begin{equation}
\mathbf{B}_{\mathrm{loop},\rho}(z,\rho, r, I) = B_0 \frac{\gamma}{\pi Q}\left[E(k)\frac{1+\alpha^2+\beta^2}{Q-4\alpha}-K(k)\right]\mathbf{\hat{\rho}}
\end{equation}
where $\alpha = \rho/r$, $\beta = z/r$, $\gamma = z/\rho$, $Q = (1+\alpha)^2+\beta^2$, $K = \sqrt{4\alpha/Q}$, $B_0 = \frac{\mu_0I}{2r}$, $K(k)$ is the complete elliptic integral function of the first kind and $E(k)$ is the one of the second kind. Given the single loop expressions, we can write the magnetic field at an arbitrary point $P_2$ in a similar fashion to Eq.~\eqref{eq:B_bobbin_axis}:
\begin{equation}
\label{eq:B_bobbin_arbitrary}
\mathbf{B}_{\mathrm{coil}}(z, \rho, r_{\mathrm{in}}, r_{\mathrm{ex}}, h, t, p, I) = \sum_{i = 1}^{N}\sum_{j = 1}^{M}\mathbf{B}_{\mathrm{loop}}(z_j, \rho, r_i,I),
\end{equation}
\begin{equation}
\mathbf{B}_{\mathrm{loop}}(z_j, \rho, r_i,I) = \mathbf{B}_{\mathrm{loop}, z}(z_j, \rho, r_i,I) +  \mathbf{B}_{\mathrm{loop}, \rho}(z_j, \rho, r_i,I).
\end{equation}

Given a coordinate system centered at the sample position, three coils are placed along the $x$, $y$, and $z$ axes at positions $\mathbf{r}_x$, $\mathbf{r}_y$, and $\mathbf{r}_z$ (Figure~\ref{fig:calculation}(d)). The polarity of each coil is determined by the direction of current flow. Each bobbin has a side with two holes for the wire to enter and exit after winding it (see Figure~\ref{fig:calculation}(c)). If the current enters the outer hole, the polarity is from the hole side to the opposite face. Using this setup, we can compute the total magnetic field at any point for given current values.

\subsection{Optimization of the parameters}

\begin{figure*}[htb!]
\centering
\includegraphics[width = 0.9 \linewidth]{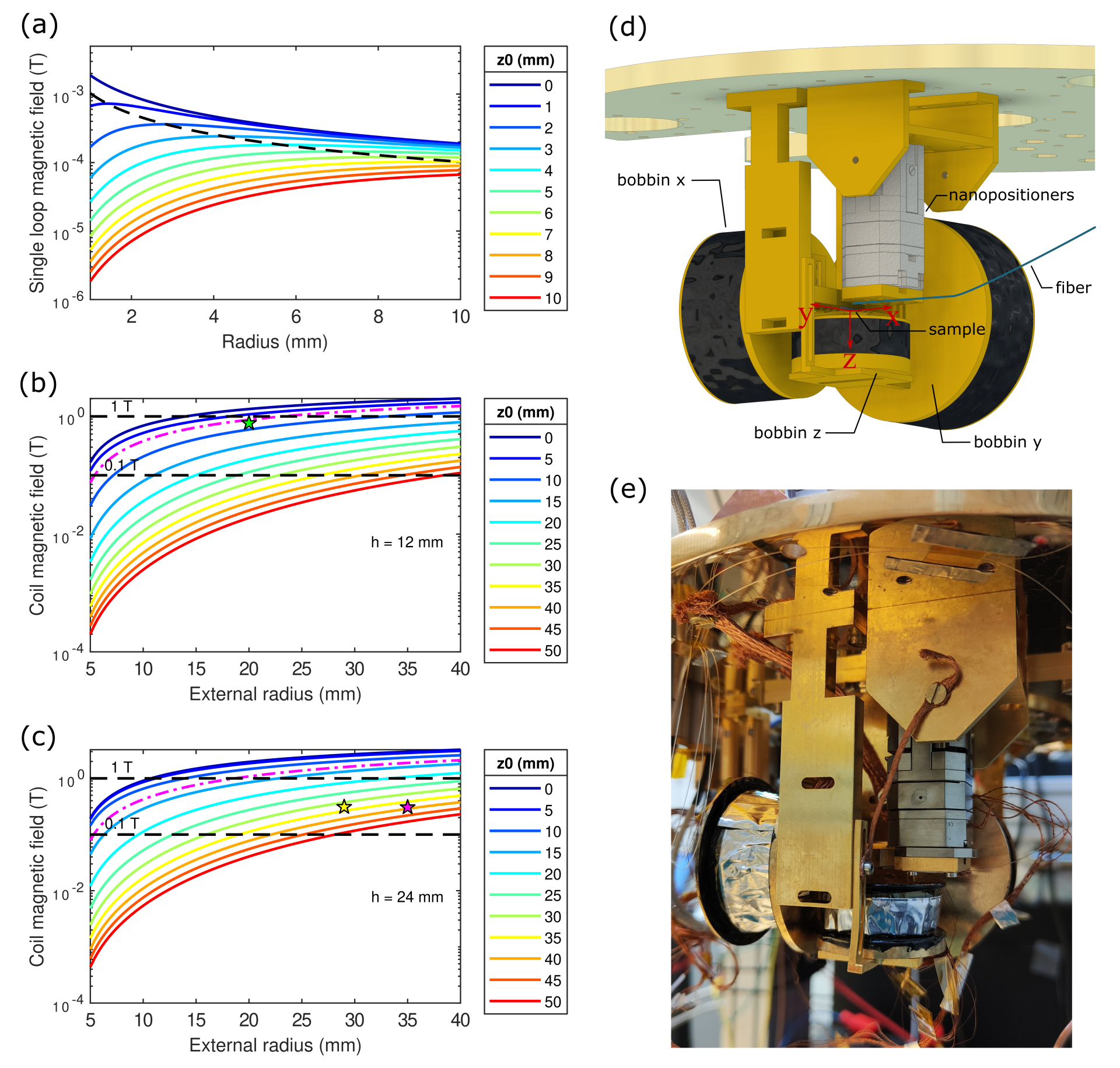}
\caption{(a)-(c) Calculated magnetic field for a current of \qty{3}{\ampere}. (a) Single loop case. On the $x-$axis we report the loop radius and different colors correspond to points on the axis with varying distance $z_0$ from the loop. The dashed black curve highlights the radii that satisfy Eq.~\eqref{eq:max_r}. (b) Coil with $h = \qty{12}{\milli\meter}$. On the $x-$axis we report $r_{\mathrm{ex}}$ and different colors correspond to points on the axis with varying $z_0$. The dashed-dotted magenta curve satisfies $z_0 = h/2$, which is the minimum allowed $z_0$. The two dashed lines indicate the condition to have magnetic field equal to \qty{0.1}{\tesla} and \qty{1}{\tesla}. The green pentagon corresponds to the bobbin $z$ design values. (c) Same as (b) but for coil with $h = \qty{24}{\milli\meter}$. The yellow and magenta pentagons correspond to the design parameters of bobbin $x$ and $y$, respectively. (d) Final design of the vector magnet showing the three bobbins, attached to the sample holder and the nanopositioners, placed on another holder. (e) Photo of the vector magnet assembly inside the DR.}
\label{fig:B_field_vs_parameters}
\end{figure*}

With the ability to calculate the expected field, we are now able to optimize the vector magnet parameters to maximize the magnetic field at the sample position while maintaining a fixed current in each coil. This optimization must consider individual bobbin properties, practical constraints, and physical limitations when assembling the system.

The wire thickness should be as small as possible to maximize the number of windings. However, excessively thin wires may be fragile and prone to breaking during the winding process. For our design, we select a wire thickness $t = 0.127$~mm. The pitch $p$ is set equal to the wire thickness $t$, ensuring compact winding. The internal radius $r_{\mathrm{in}}$ is also chosen to be as small as possible, constrained at 4 mm to avoid excessive deformation of the copper layer. The maximum bobbin height is limited by the winding machine stage ($\sim$~45~mm) and the available space in the DR. The key adjustable parameters are the external radius $r_{\mathrm{ex}}$ and the distance $z$, with $r_{\mathrm{ex}}$ constrained to a maximum of approximately 40~mm by the winding machine. For clarity, we define $z_0 = z-h/2$, representing the distance of a given point from the bobbin center.

For a single loop, an optimal radius $r_{\mathrm{max}}$ exists for a given distance $z_0$ (see dashed black curve in Figure~\ref{fig:B_field_vs_parameters}(a)):
\begin{equation}
\label{eq:max_r}
r_{\mathrm{max}} = \sqrt{2} z_0.
\end{equation}
However, this relationship does not hold for a coil, as the total field results from the sum of multiple loops (Eq.~\eqref{eq:B_bobbin_axis}). In this case, increasing $r_{\mathrm{ex}}$ while keeping other parameters constant always results in a stronger field. Thus, to maximize the magnetic field, the bobbin should be as close as possible to the sample while having the largest feasible $r_{\mathrm{ex}}$ (Figure~\ref{fig:B_field_vs_parameters}(b)).

For a single coil, an ideal configuration is to make the internal radius large enough to accommodate the sample at $z_0 = 0$. However, in the case of a three-axis vector magnet with optical access via a lensed fiber, this is not feasible. Instead, the minimum achievable distance $z_0$ is constrained by half of the bobbin height.

It is also impractical to make all bobbins arbitrarily large and close to the sample, as they would physically interfere with one another. Our goal is to design coils capable of generating at least 100~mT at 3~A -- a reasonable current for standard current sources -- and as close to 1~T as possible. Figure~\ref{fig:B_field_vs_parameters}(b) shows the calculated magnetic field as a function of $r_{\mathrm{ex}}$ for various values of $z_0$. The current is 3~A and the height is fixed to the value chosen for bobbin $z$, which is 12~mm. The black dashed lines mark the limits of 0.1~T or 1~T, with the area in between the lines representing the allowed parameter space.

In general, minimizing $z_0$ is preferable to increasing $r_{\mathrm{ex}}$, as smaller bobbins offer several advantages:\ they are simpler to assemble, occupy less space in the DR, are lighter and easier to thermalize, and are expected to exhibit more stable superconductivity. The dashed-dotted magenta curve in Figure~\ref{fig:B_field_vs_parameters}(b) represents $z_0 = h/2$, the smallest allowable $z_0$, with points below this curve falling within the feasible design space.

Due to spatial constraints, three coils cannot simultaneously reach their optimal setting. We set the $z-$axis as the principal one and choose the parameters labelled as the green pentagon in Figure~\ref{fig:B_field_vs_parameters}(b) for bobbin $z$. The resulting field at the sample position is close to 1~T. The parameters for bobbins $x$ and $y$ are then limited by the design of bobbin $z$. The final choices, represented by the yellow and magenta pentagons in Figure~\ref{fig:B_field_vs_parameters}(c), yield fields of approximately 0.3~T at the sample position.
Given all these constraints, we pick our final design for a vector magnet (shown in Figures~\ref{fig:B_field_vs_parameters}(d-e)). The selected design parameters for each bobbin are summarized in Table~\ref{tab:design}. We highlight that when mounting the bobbins on the sample holder, we have some degrees of freedom to let us align their axes with the center of the sample. Furthermore, bobbin $z$ distance from the sample is tunable and, ideally, it can be placed as close as desired to the sample, with the only limitation given by the nanopositioners stack, which results in a minimum distance $z_0 = 10.5$ mm.

\begin{table*}
\caption{\label{tab:design}Bobbin design parameters. The magnetic field at the sample position and in the center of the coil when a current of \qty{3}{\ampere} is applied are also shown.}
\begin{ruledtabular}
\begin{tabular}{ccccccccc}
Bobbin&$r_{\mathrm{in}}$ (mm)&$r_{\mathrm{ex}}$ (mm)&$h$ (mm)&$t$ (mm)&$p$ (mm)&$z_0$ (mm)& $B$ at sample (mT) & $B$ in center (T)\\
\hline
$x$ & 4& 29& 24& 0.127& 0.127& 37.5& 309& 3.59\\
$y$ & 4& 35& 24& 0.127& 0.127& 45.0&  305& 4.09\\
$z$ & 4& 20& 12& 0.127& 0.127& 10.5&  767& 1.80
\end{tabular}
\end{ruledtabular}
\end{table*}

\section{Vector magnet assembly}
\label{sec:assembling}

This section gives a detailed description on how to make a bobbin and to then estimate its effective parameters. Furthermore, we will describe how assemble the vector magnet inside the DR in order to ensure optimal thermalization.

\subsection{Winding procedure}
\label{sec:winding}

\begin{figure}[htb!]
\centering
\includegraphics[width = 0.9 \linewidth]{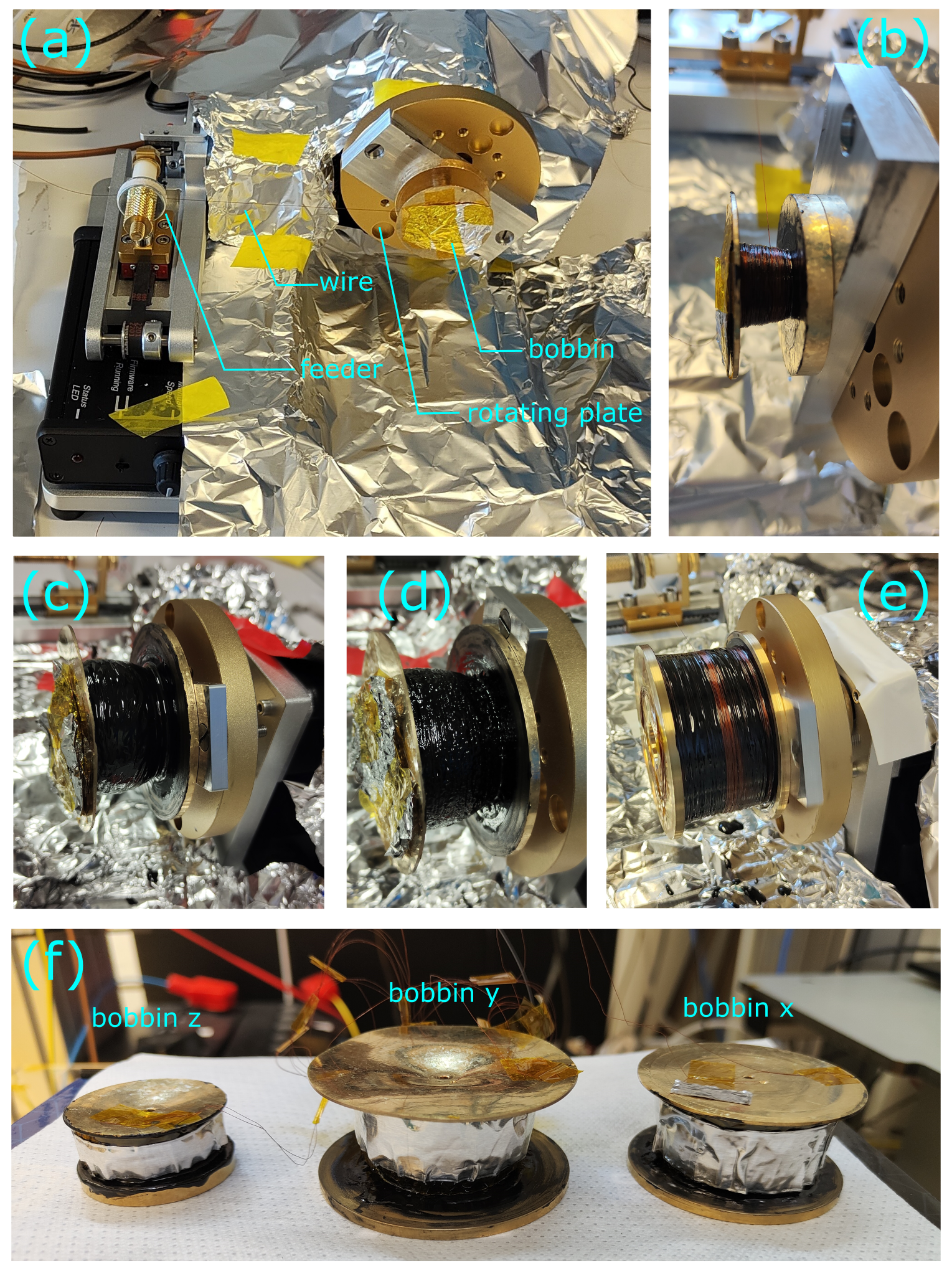}
\caption{Winding process. (a) Picture of the winding machine. (b) Detail of a bobbin during the winding process. (c) Picture of a bobbin after completed winding. There is a noticeable wire accumulation on one side. (d) Picture of the bobbin after degassing. The surface roughness indicates the removal of air bubbles. (e) Picture of an adjusted bobbin design with thicker side plate, making the wire distribution more uniform. (f) Picture of the three completed bobbins. Note that bobbin $z$ and bobbin $y$ have a non flat side plate.}
\label{fig:winding}
\end{figure}

The bobbins are made out of copper, then annealed in vacuum at \qty{840}{\celsius} for 48~hours, followed by gold plating.  Annealing improves thermalization, while gold plating prevents oxidation. For winding, we use Supercon Cu:SC wires (model T48B-M), where SC denotes the superconducting material - in this case, NbTi. The bare wire thickness is \qty{0.100}{mm}, and it is coated with a FORMVAR insulation layer, resulting in a total thickness of \qty{0.127}{mm}.

We wind the bobbins using a semi-automatic coil winder (CNC Mini Coil Winder MK5) (see Figures~\ref{fig:winding}(a-b)). Before winding the Cu:SC wire, we calibrate the machine using a low-cost copper wire of similar thickness. This involves winding a few layers, adjusting the parameters, unwinding, and repeating the process until achieving reproducible and satisfactory results. Depending on the bobbin dimensions, the entire winding process takes several hours.

After winding each layer, we apply Stycast 2850 epoxy with catalyst CAT 24LV or 23LV (which has a longer curing time)~\cite{Raha2021}. The epoxy encapsulates the coil, preventing movement under Lorentz forces and reducing the likelihood of quenching. Additionally, Stycast is thermally conductive, aiding heat dissipation in the event of superconductivity loss, and electrically insulating, improving isolation between loops. To ensure even epoxy distribution, we continuously rotate the bobbin on the winding machine, avoiding prolonged stops that could cause epoxy accumulation due to gravity. Periodically, we check for electrical shorts between the gold-plated bobbin and the wire ends, as damage to the thin insulation layer could lead to unintended contact. We also monitor whether the resistance between the wire ends is constant over time.

Once the bobbin is fully wound, we cut the wire and continue spinning for one or two hours (depending on the curing agent) to allow the epoxy to partially solidify. We then degas the bobbin in a vacuum chamber at pressures below \qty{1}{mbar} for about 10 minutes, ensuring the removal of air bubbles before complete epoxy solidification. Figures~\ref{fig:winding}(c-d) show the bobbin before and after degassing, clearly illustrating the expulsion of trapped air, denoted by the surface roughness change.

After degassing, we keep the bobbin spinning on the machine for another 12 hours and then allow an additional 1--2 days for complete drying, including the inner layers. To prevent contamination in the DR, we perform a second degassing step for at least 1 hour. The final bobbins appear as shown in Figure~\ref{fig:winding}(f).

During winding, we observe wire accumulation on one side of the bobbin. Additionally, after degassing, the bobbin plate on that side is no longer flat (cf.\ Figures~\ref{fig:winding}(c), (d), and (f)). We identify these issues as being caused by the use of a thin (\qty{1}{mm}) bobbin plate, which becomes excessively soft after annealing. By using a thicker plate, we significantly reduce these deformations, as shown in Figure~\ref{fig:winding}(e). Another challenge we encounter is the wire breakage during winding. In some instances, epoxy residue accumulates in the wire feeder, becoming stickier over time and eventually severing the wire after around 1--2 hours. In other cases, the wire breaks due to too high tension. For the bobbin shown in Figure~\ref{fig:winding}(e) we improve the way we feed the wire and reduce the tension on it. We manage to wind $\approx 32$ thousands windings, i.e. the full frame, without  breaking the wire. If needed, in the future, we could further mitigate this issue by using a thicker wire and/or a proper tensioner. If the wire breaks despite these measures, one possible solution is to fully dry the epoxy, as described above, and then wind additional wire to be soldered to the original one. However, this approach has not been tested and may impact the superconductivity of the coil. Even though now we are able to make bobbins avoiding the issues encountered in the winding process, in the following we will refer to the bobbins shown in Figure~\ref{fig:winding}(f), since those have been made to be used in the vector magnet setup (Figures~\ref{fig:B_field_vs_parameters}(d-f)). 

\subsection{Estimation of the effective parameters}
\label{sec:effective_parameters}
Since the wound bobbins are neither full nor homogeneous, we need to adjust the actual coil parameters to account for these imperfections to be able to estimate the generated magnetic field accurately. The actual height $h$ is reduced by 0.5--1~mm as the wire cannot be perfectly aligned with the side walls. The pitch $p$ is set in the machine to \qty{0.15}{mm}, while the internal radius $r_{\mathrm{in}}$ is kept as designed. As the thickness $t$ increases due to the epoxy between the layers, and the external radius $r_{\mathrm{ex}}$ is not uniform because of winding irregularities, these parameters are the most challenging to determine. We therefore define effective values such that the total magnetic field generated by this coil matches the actual one.

Several measurable quantities help us estimate these parameters (see Table~\ref{tab:measured_bobbins}). The machine logs provide the total number of windings and layers and by measuring the wire resistance, we estimate its length. Both the effective thickness and the external radius influence these values. We measure the minimum and maximum $r_{\mathrm{ex}}$ and select an effective value within this range.

\begin{table*}
\caption{\label{tab:measured_bobbins}Measured bobbin parameters. The wire length is inferred from the measured resistance. The deformation is defined as the distance between the center (original undeformed plate) and the point of maximum deformation.}
\begin{ruledtabular}
\begin{tabular}{ccccccccc}
Bobbin&Windings&Layers&Resistance (k$\Omega$)&Wire length (km)&Weight (g)&Min $r_{\mathrm{ex}}$ (mm)& Max $r_{\mathrm{ex}}$ (mm)& Deformation (mm)\\
\hline
 $x$& 19126          & 122& 5.88& 1.65& 233.3& 20.85& 24.25& 0.7\\
 $y$& $\approx 23000$& 150& 7.62& 2.13& 300.0& 20.25& 23.5& 4.6\\
 $z$& 7240&             94& 1.82& 0.51& 116.0& 17.25& 18.6& 1.5\\
\end{tabular}
\end{ruledtabular}
\end{table*}

With these considerations, we estimate the effective parameters, which are listed in Table~\ref{tab:effective_bobbins}, along with the expected magnetic field at the sample position and in the center of each bobbin. The measured values for the number of windings, layers, and wire length agree with those derived from the effective parameters. For bobbin $y$ the plate deformation, \qty{4.6}{mm}, is particularly significant. This means that the height is also not homogeneous, thus the effective external radius must take this into account. Due to the epoxy layer and the reduced external radius, the expected magnetic field is significantly lower than in the original design -- by about a factor of 3.3 for bobbin $y$, 2.7 for bobbin $x$, and 1.9 for bobbin $z$.

\begin{table*}
\caption{\label{tab:effective_bobbins}Effective bobbin parameters. Number of windings, layers and the wire length estimated from these parameters. The magnetic field at the sample position and in the center of the coil when a current of \qty{3}{\ampere} is applied are also shown.}
\begin{ruledtabular}
\begin{tabular}{cccccccccccc}
Bobbin& Windings& Layers& Wire length (km)& $r_{\mathrm{in}}$ (mm)&$r_{\mathrm{ex}}$ (mm)&$h$ (mm)&$t$ (mm)&$p$ (mm)&$z_0$ (mm)& $B$ at sample (mT) & $B$ in center (T)\\
\hline
$x$ & 19188& 123& 1.55& 4& 21.7& 23.5& 0.143& 0.15& 37.5& 115& 2.13\\
$y$ & 22950& 150& 1.99& 4& 23.5& 23.0& 0.130& 0.15& 45.0& 93& 2.49\\
$z$ & 7524&   99& 0.52& 4& 18.0& 11.5& 0.140& 0.15& 11.9& 397& 1.23 
\end{tabular}
\end{ruledtabular}
\end{table*}

\begin{figure*}[htb!]
	\centering
	\includegraphics[width = 0.9 \linewidth]{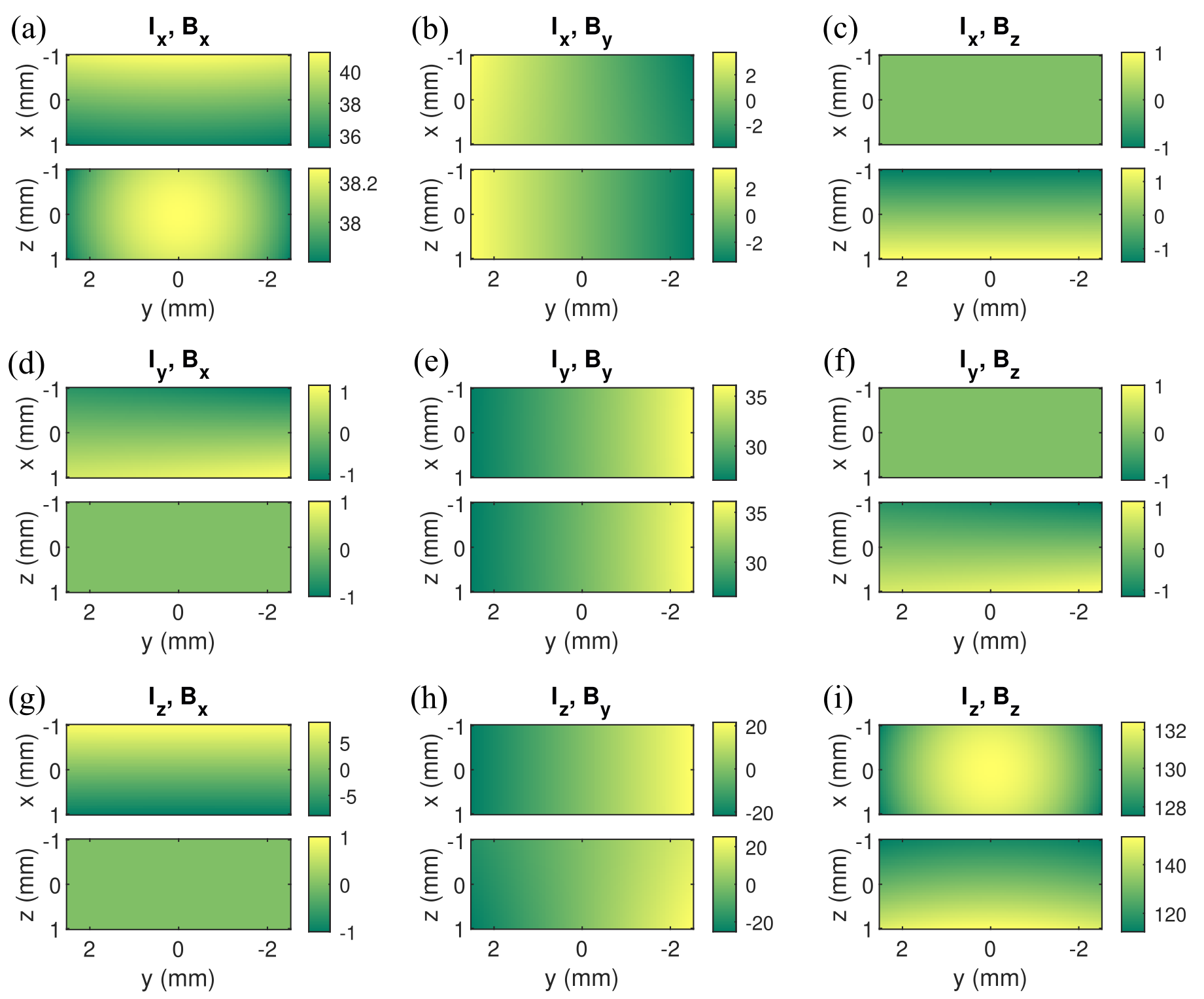}
	\caption{Colormaps of the calculated magnetic field (in \qty{}{\milli\tesla}) when applying \qty{1}{\ampere} to a coil, modeled with the effective parameters. We sweep the coordinates around the vector magnet origin, corresponding to the center of the sample. The titles indicate the bobbin to which we apply the current and which magnetic field component is considered.}
	\label{fig:B_field_maps}
\end{figure*}

While the calculated magnetic field values correspond to the exact center of the sample, the actual nanophotonics devices are distributed over a \qty{5}{mm} range along the $y-$axis. Additionally, potential misalignment during mounting must also be considered. It is therefore useful to estimate the magnetic field distribution over the chip surface to assess deviations from the ideal case. Figure~\ref{fig:B_field_maps} shows calculated magnetic field colormaps in the plane of the chip's top surface ($xy$ plane) and along the wide side of the chip ($yz$ plane). Each plot represents the magnetic field $x$, $y$, or $z$ component when applying a \qty{1}{\ampere} current to bobbin $x$, $y$, or $z$. The range in the $y-$direction corresponds to the nanopositioner range, meaning that we can reach devices in that region. For the $x-$ and $z-$axes, we use a \qty{2}{\milli\meter} range to account for potential misalignment. The magnetic field component along the bobbin axis (Figures~\ref{fig:B_field_maps}(a), (e), and (i)) varies by approximately $10\%$ within the considered planes. As discussed in Section~\ref{sec:mag_field_meas}, this variation is comparable to the uncertainty of the calculated field. The off-axis field components are negligible for bobbins $x$ and $y$, but not for bobbin $z$ (cf.\ Figure~\ref{fig:B_field_maps}). In particular, when measuring devices far from the chip center, the $y-$component of the magnetic field generated by bobbin $z$ is comparable to that generated by bobbin $y$. This effect may need to be considered when precise magnetic field calibration is required.

\subsection{Thermalization in the dilution refrigerator}
\label{sec:wiring} 

\begin{figure*}[htb!]
\centering
\includegraphics[width = 0.9 \linewidth]{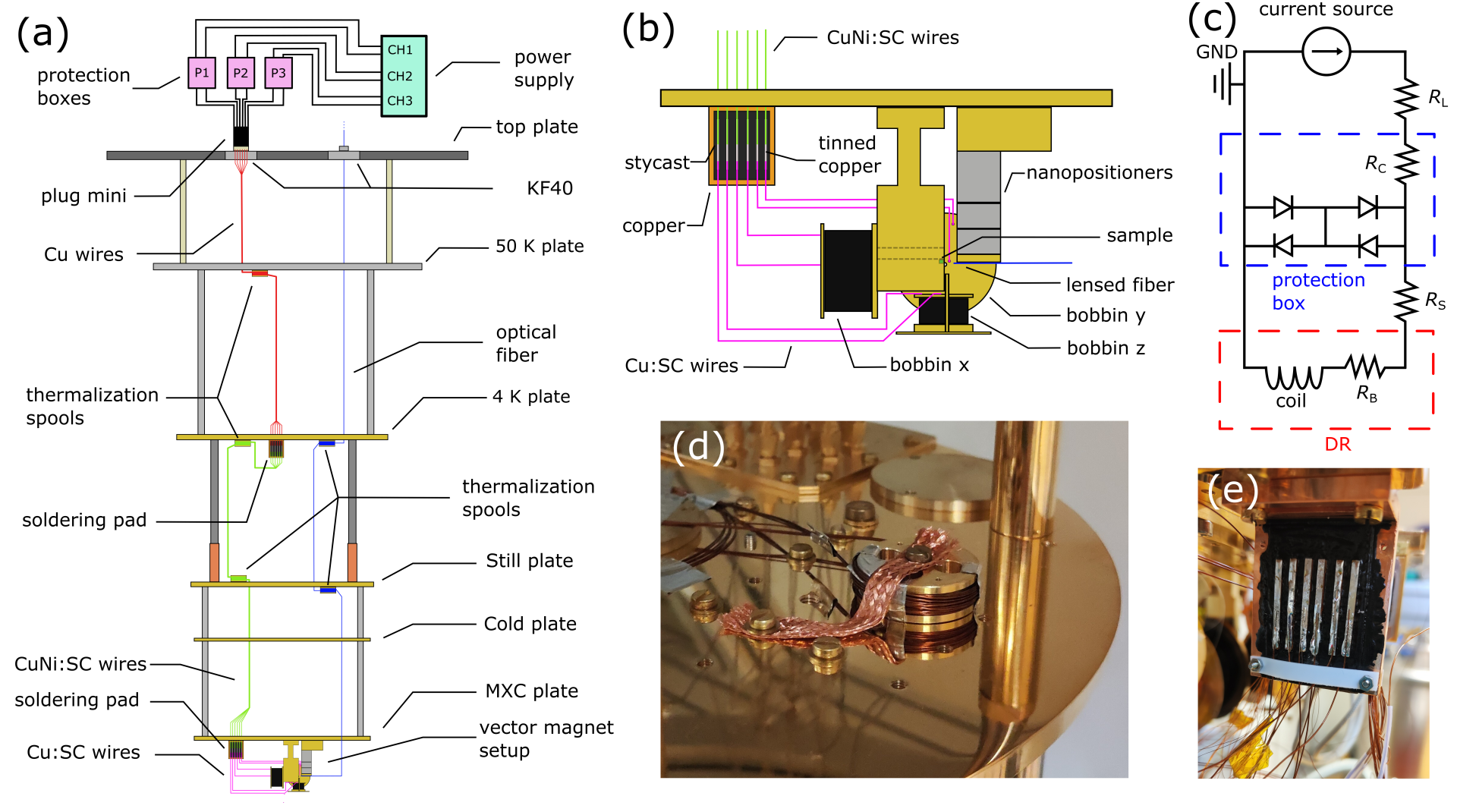}
\caption{Scheme of the full vector magnet setup. (a) Schematic of the dilution refrigerator, including wirings. (b) Details of the MXC plate. (c) Schematic of the electrical connections, including the protection circuit. (d) Picture of a thermalization spool. (e) Picture of a soldering pad.}
\label{fig:wiring}
\end{figure*}

The vector magnet is mounted on the MXC plate of a Bluefors DR (model LD250), which reaches a base temperature of approximately \qty{10}{\milli\kelvin} before installation. Proper thermalization of both the bobbins and the wires connected to the outside is essential to reduce the impact on the achievable base temperature. We aim to minimize wire resistance while ensuring good thermal contact at each flange. In the event of a quench -- a sudden loss of superconductivity -- it is however crucial to prevent heat propagation, which could further spread the quench and affect other parts of the wiring and the overall performance of the DR. The wiring scheme depicted in Figure~\ref{fig:wiring}(a-b) takes these requirements into account.

The bobbin frames are anchored to the MXC plate via the sample holder and copper braids, promoting coil superconductivity. As described in Section~\ref{sec:winding}, the coils are made of \qty{0.1}{\milli\meter} thick Cu:SC wire encapsulated in Stycast, which aids heat dissipation in case of a quench. The wires are soldered to thicker (\qty{0.4}{\milli\meter}) CuNi:SC wires (model SW-18). Notably, we use CuNi instead of Cu due to its lower thermal conductivity, which helps prevent heat propagation between different flanges of the DR. These CuNi:SC wires extend up to the \qty{4}{\kelvin} stage, where they are soldered to annealed \qty{0.4}{\milli\meter} Cu wires. This transition is necessary because the CuNi:SC wires would introduce excessive resistance above their superconducting transition temperature at approximately \qty{9}{\kelvin}. The Cu wires are soldered to a 10-pin plug mini connector inserted into a feedthrough at the top of the DR. Along the DR, we use annealed and gold-plated copper spools to improve thermalization (see Figure~\ref{fig:wiring}(d)). The wires are wound around each of the spool for five turns. One spool is installed at \qty{50}{\kelvin} for the copper wires, and two spools -- one at the  \qty{4}{\kelvin} plate and one at the Still plate (\qty{1}{\kelvin}) -- are installed for the CuNi:SC wires.

The soldering points at the \qty{4}{\kelvin} and MXC plates must be both thermally anchored and electrically isolated. To achieve this, we design soldering pads (see Figure~\ref{fig:wiring}(e)). These are copper pads screwed to the DR plate for thermal anchoring, with pre-tinned copper strips glued on top using Stycast 2850. Each strip accommodates two wires for soldering. Since the epoxy is thermally conductive but electrically insulating, it preserves thermal contact while preventing electrical shorts to the DR ground. To reinforce the electrical isolation, we first glue a thin paper sheet onto the pad before applying the epoxy and copper strips.

After installing the vector magnet, the DR base temperature increases slightly from around \qty{10}{\milli\kelvin} to \qty{12}{\milli\kelvin}, before applying any current. Most of our tests, however, are conducted at \qty{27}{\milli\kelvin}, as a later unrelated additional component to the DR further raised the MXC plate temperature. During current ramp-up, the MXC plate temperature can temporarily reach up to around \qty{200}{\milli\kelvin} due to the inductance of the bobbins. However, once the target current (max.\ \qty{3}{\ampere}) stabilizes, the temperature returns to within a few \qty{}{\milli\kelvin} of the base temperature. For instance, starting from a base temperature around \qty{13}{\milli\kelvin} at zero current, when all the three bobbins operate simultaneously at \qty{3}{\ampere}, the MXC plate temperature rises to \qty{23}{\milli\kelvin}.

\section{Safety measures}
\label{sec:safety}

During the current ramp-up of the coils quenching occasionally occurs, leading to a sudden increase in the DR temperature. To mitigate these heating effects from quenching, we have implemented preventive measures~\cite{Smith1963,Maddock1968}.

\subsection{Protection diodes}

\begin{figure}[htb!]
\centering
\includegraphics[width = 0.9 \linewidth]{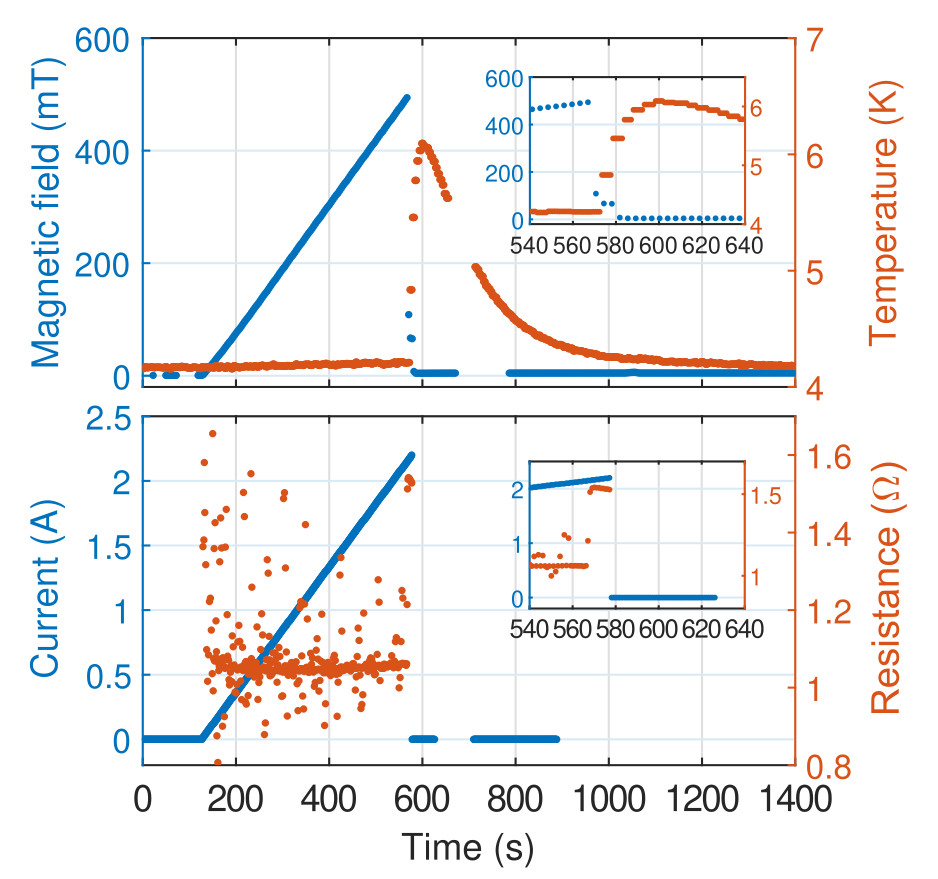}
\caption{Training of bobbin $x$ (positioned at (-37.5, 0, 0)~mm) consisting of a current ramp-up and controlled quench event, with all safety measures described in the main text employed. In the top panel, the $x$ component of the magnetic field, measured with a Hall probe at position ($-21.3$, 0, $-8.44$)~mm, as a function of time is shown in blue. The temperature of the MXC plate (when only the pulsed tube is used) in red. In the bottom panel the applied current and the total resistance $R$ are shown in blue and red, respectively, (see Appendix~\ref{app:protection_circuit}) measured with the current source. In both panels, the insets show the quench event in more detail.}
\label{fig:quench}
\end{figure}

In order to prevent excessive heating during a quench, we place a bridge circuit in parallel with the coils. If the voltage drop in the DR exceeds the diode threshold -- which is the case when the superconductivity is lost -- the current preferentially flows through the diodes instead of the coil, limiting further heating (see Figure~\ref{fig:wiring}(c)). Additionally, this circuit also acts as a rectifier, filtering AC components of the current. We position the circuit outside the DR, as close as possible to the feedthrough. Placing it inside the DR at the \qty{4}{\kelvin} or Still plate would improve heat dissipation further, however, would require cryogenic components which we did not implement. Due to the cable resistance outside the DR ($R_\mathrm{S}$ in Figure~\ref{fig:wiring}(c)), the maximum allowable current before reaching the diode threshold voltage is therefore \qty{17.5}{\ampere} (see calculation in Appendix~\ref{app:protection_circuit}).

During a quench, we observe the bobbin resistance rising from 0 to \qty{0.44}{\ohm} within a few seconds. If the current source is not stopped, the resistance continues to increase gradually, despite this protection. In parallel, the DR base temperature quickly climbs from \qty{20}{\milli\kelvin} to \qty{1}{\kelvin} in dilution refrigeration mode (and from \qty{4}{\kelvin} to \qty{6}{\kelvin} when only the pulse tube is used, as shown in Figure~\ref{fig:quench}), and then continues to rise. In fact, the residual circulating current in the coil dissipates power on the order of Watts (see Appendix~\ref{app:protection_circuit}), which explains the continued heating. This indicates that the protection circuit alone is insufficient.

\subsection{Voltage limit and temperature check}
Our second safety measure is a voltage limit in the power supply. We set this limit above the required operating voltage but below the quench voltage. This ensures that if a quench occurs, the power supply shuts off automatically. When setting a specific current value we must gradually ramp up to the desired value, as otherwise heating occurs due to the induced electromotive force, since the coils behave as inductors. To mitigate this effect, we typically increase the current in steps of \qty{50}{\milli\ampere} per second, adjusting the voltage limit accordingly at each step.

We observe that at low current values, the voltage limit is sometimes reached even in the absence of a quench. We attribute this to voltage fluctuations during ramp-up and imprecise voltage and current measurements at small values. This effect is visible in Figure~\ref{fig:quench}, where the resistance exhibits large fluctuations. To address this, we set a higher voltage limit for current values below \qty{0.5}{\ampere}. However, in some cases, the issue persists, requiring an increased limit even for higher currents. Once the target current is reached, we can safely lower the voltage limit. As a consequence, the coils do not operate under fully protected conditions during ramp-up. To address this, we implement a script that monitors the MXC plate temperature. If the temperature exceeds a predefined threshold (\qty{400}{\milli\kelvin} in dilution refrigeration mode, \qty{5}{\kelvin} when only the pulse tube is used), the current source is shut off.

\subsection{Training}
The coils, especially those with more windings, require training -- repeated ramps in the current -- to reach high current values~\cite{Raha2021}. During each cooldown, when we increase the current for the first time, the coils typically quench at around \qty{2}{\ampere}. Our safety measures allow us to handle this safely, although the DR temperature rises rapidly -- from around \qty{100}{\milli\kelvin} to \qty{1}{\kelvin} in dilution refrigeration mode and from \qty{4}{\kelvin} to \qty{6}{\kelvin} when only the pulse tube is used. After a quench, we wait for at least half an hour for the temperature to drop, ensuring that the coil is superconducting again. The ramp-up process is then repeated, and the quench current increases with each cycle. To reach \qty{3}{\ampere}, we typically require between one to three training cycles. While our current source is limited to \qty{3}{\ampere}, larger currents should in principle be possible.

The quenching occurs due to increasing magnetic forces during the current ramp-up, straining the wires, despite them being embedded in epoxy. After quenching, the bobbin settles into a more stable configuration, allowing it to sustain higher currents in subsequent cycles. We also observe that when running multiple bobbins simultaneously, they must be trained together. The combined magnetic forces are stronger and act in different directions, necessitating additional training to maintain stability.

\section{Magnetic field measurements}
\label{sec:mag_field_meas}

In order to measure the magnetic field generated by the coils and verify the validity of our model we use 2D Hall probes (HHP-NP Arepoc). As a first step, we confirm that the magnetic field increases linearly with the current (see Figure~\ref{fig:quench}). In total we use three sensors, each measuring the magnetic field component along one of the bobbin axes. The magnetic field of primary interest is the one at the sample position, however, placing the Hall probes exactly at the sample location is not feasible. For each sensor, we calculate the expected magnetic field when a given current flows through the corresponding coil and assume that the actual magnetic field at the sample scales with the expected field in the same way that the Hall probe measurement scales with the expected value. Table~\ref{tab:hall_meas} summarizes the measured and expected values. We attribute the observed discrepancies of up to $11.3\%$ to the uncertainty in the Hall sensor sensitivity and positioning -- mainly due to thermal contraction -- geometrical deviations of the bobbins from the ideal model, and external sources of magnetic noise.

\begin{table}
\caption{\label{tab:hall_meas} Measured and calculated magnetic field with fixed Hall probes. In each row, we indicate to which coil we apply \qty{1}{\ampere} and which Hall sensor is used. In the considered reference system, centered at the sample, the positions of the center of bobbin $x$, $y$ and $z$ are ($-37.5$, 0, 0)~mm, (0, $-45.0$, 0)~mm, and (0, 0, 11.9)~mm, respectively. Hall sensor $x$, $y$, and $z$ positions are ($-21.3$, 0, $-8.44$)~mm, (2.24, $-28.8$, $-0.7$)~mm, and (2.24, 9.55, 0.5)~mm, respectively. The relative discrepancy between the measured and expected magnetic fields is also shown.}
\begin{ruledtabular}
\begin{tabular}{ccccc}
Bobbin&Hall sensor&$B_{\mathrm{meas}}$ (mT)&$B_{\mathrm{exp}}$ (mT)&Discrepancy (\%)\\
\hline
$x$ & $x$& 68.24& 61.33& $+11.3$\\
$y$ & $y$& 924.0& 959.2& $-3.7$\\
$z$ & $z$&  242.9& 244.0& $-0.4$
\end{tabular}
\end{ruledtabular}
\end{table}

\begin{figure*}[htb!]
\centering
\includegraphics[width = 0.9 \linewidth]{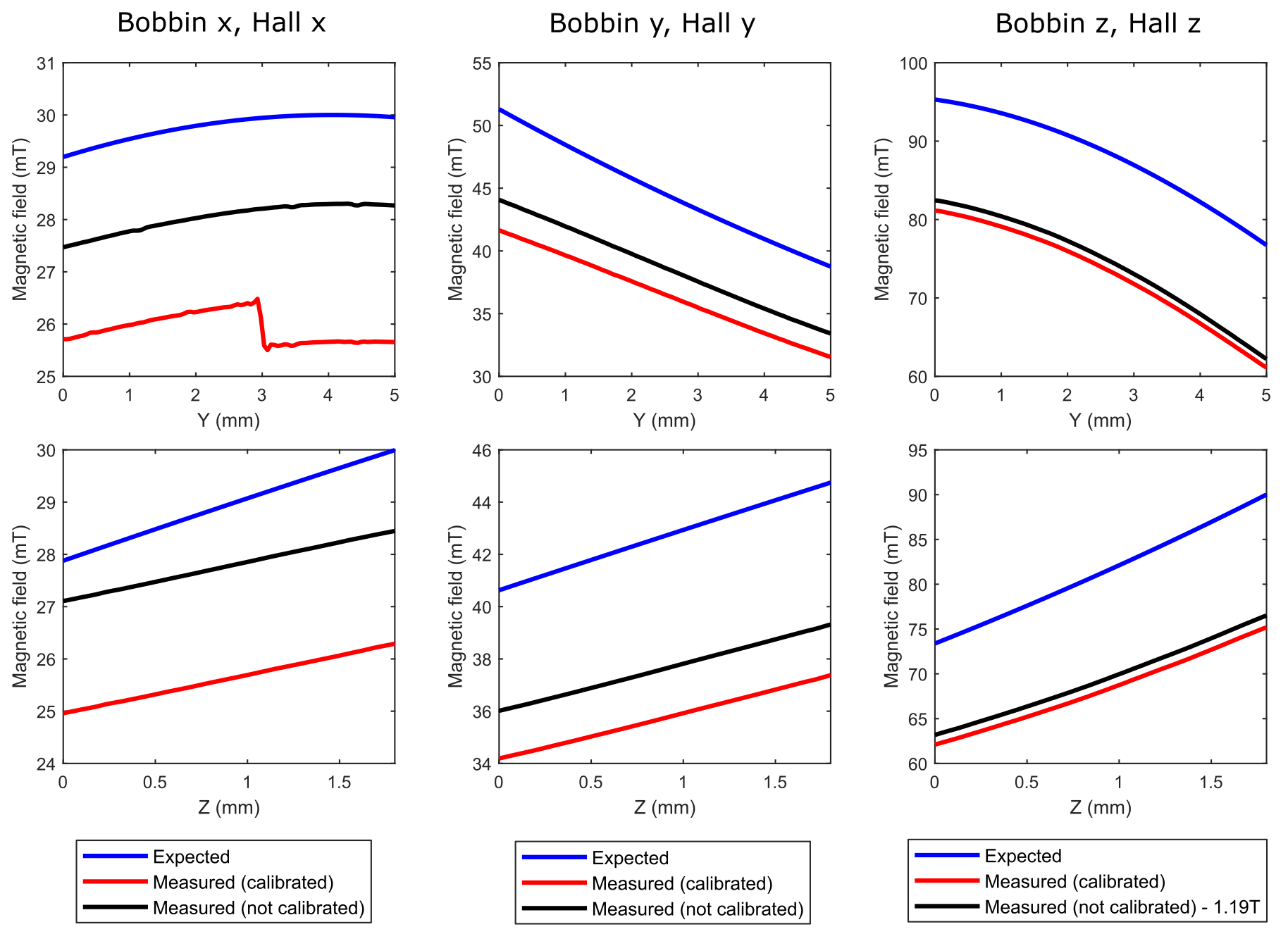}
\caption{Magnetic field measurements when sweeping the Hall sensors positions. In the considered reference system, centered at the sample, the positions of the center of bobbin $x$, $y$ and $z$ are ($-37.5$, 0, 0)~mm, (0, $-45.0$, 0)~mm, and (0, 0, 12.75)~mm, respectively. Hall sensor $x$, $y$, and $z$ positions are ($0.3$, $-4.06+Y$, $-13.25 + Z$)~mm, ($10.74$, $-15.70 + Y$, $-14.55 + Z$)~mm, and ($8.14$, $1.05 + Y$, $-0.5 + Z$)~mm, respectively. Here, $Y$ and $Z$ are the nanopositioner coordinates, which can vary between $0$ and $5$~mm.
The positions of the sensors are varied in the $y$ (upper plots) and $z$ (lower plots) directions. In the former case, $Z = 1.7$~mm, in the latter $Y = 2.5$~mm. The magnetic field is measured with \qty{1}{\ampere} applied to the coil indicated in the title using the indicated Hall sensor. Blue trace:\ calculated magnetic field. Black trace:\ measured magnetic field, before calibration. Red trace:\ measured magnetic field subtracted with the zero current calibration. For Hall probe $z$ measurements, we subtract \qty{1.19}{\tesla} from the black trace to  make the plot more readable.}
\label{fig:attocube_hall}
\end{figure*}

In this configuration, we measure the magnetic field at a fixed position, where the sensors are placed. To obtain an even more detailed insight into the behavior of the magnets, we proceed to mount the Hall probes on nanopositioners, allowing us to scan the sensor position along different axes. First, we perform a scan without applying any current to the coils, enabling us to account for intrinsic sensor offsets and environmental magnetic sources. Additionally, we test the system by ramping the current up to \qty{3}{\ampere} and back down to zero several times. We observe that the measured magnetic field at zero current varies by a few \qty{}{mT} and it drifts over time, potentially due to residual magnetization in the coils. Figure~\ref{fig:attocube_hall} presents the measurements obtained from scanning the nanopositioners along the $y$ and $z$ directions. For each coil, we consider a Hall probe aligned with the bobbin axis and measure the magnetic field when \qty{1}{\ampere} flows through the coil. The plot shows the measured field, the calibrated field (i.e.\ after subtracting the field measured at \qty{0}{\ampere}), and the expected field from our model.

Our results show that the measured and expected curves follow the same trend, though the absolute values typically differ by approximately $10\%$ for bobbin $x$ and even more for bobbins $y$ and $z$. The discrepancy is likely higher for these two bobbins due to their magnetic field being highly sensitive to the exact position of the nanopositioner along the $x$-axis, which has a relatively large error associated with it. In the upper plot of bobbin $x$ and Hall probe $x$, we see a jump in the calibrated curve, due to stray magnetic fields or Hall probe malfunction during the calibration measurement. Additionally, for bobbin $z$, we observe a large offset of more than \qty{1}{\tesla} at zero current, which we attribute to the Hall sensor itself. The discrepancies between expected and measured values arise from the same factors discussed earlier -- sensor uncertainty, geometric deviations, and magnetic noise. Notably, in this specific measurement, the Hall sensor positions are estimated only roughly, relying primarily on the position readings of the nanopositioners. We conclude that our model provides a good estimation of the magnetic field generated by the coils. However, the actual magnetic field in the environment can be predicted with a $\sim$\qty{10}{\milli\tesla} resolution and we attribute to it a relative error of about $10\%$, that can be higher if the probe position is not known with good accuracy, as in the nanopositioner scans. From the measurements in Table~\ref{tab:hall_meas} -- where the position of the sensor is known with a better accuracy -- we can conclude that the expected magnetic fields at the sample position and in the center of the bobbins predicted in Table~\ref{tab:effective_bobbins} are reasonable.

\section{Conclusions and outlook}
\label{sec:Conclusion}

We have provided a comprehensive manual on the design, construction, and optimization of a superconducting three-axis vector magnet. By implementing an effective thermalization strategy -- including optimized wiring in a dilution refrigeration and carefully anchored components -- we are able to individually drive each coil with \qty{3}{\ampere} while consistently maintaining a base temperature of approximately \qty{15}{\milli\kelvin}, that can rise up to \qty{23}{\milli\kelvin} when the coils operate simultaneously. The system achieves a maximum magnetic field of around \qty{2.5}{\tesla} in the center of bobbin $y$, while at the sample position we can reach \qty{0.4}{\tesla} along the $z$ axis and \qty{0.1}{\tesla} along the $x$ and $y$ axes. The limitations in field strength could be addressed by improving the winding process. Based on our design, we estimate that fully wound bobbins -- which are feasible, as demonstrated in Figure~\ref{fig:winding}(e) -- could generate magnetic fields 2 to 3 times higher at the sample position than for these initial attempts. Additionally, applying higher currents -- which has been shown for similar systems~\cite{Raha2021} -- or increasing the size of the bobbins could further increase the magnetic field. For instance, we consider bobbins with the same design parameters as in Table~\ref{tab:design}, but with $h = \qty{40}{\milli\meter}$ (the limit of the winding machine). We assume they are fully wound and that the epoxy thickness is the same as what we experienced. If we drive each coil with \qty{10}{\ampere} -- well below the protection diode threshold current (see Appendix \ref{app:protection_circuit}) -- we expect to have a magnetic field at the sample position of \qty{0.94}{\tesla}, \qty{1.06}{\tesla}, and \qty{2.28}{\tesla} for bobbin $x$, $y$, and $z$, respectively. However, such heavy magnets, driven with high currents, might lead to thermalization and quenching issues.\\
For applications requiring a strong field along a single axis, placing the sample inside the bobbin, where the field is at its maximum, is also a viable alternative. We realized this with the bobbin shown in Figure~\ref{fig:winding}(e), which has $r_{\mathrm{in}} = 8$~mm -- enough to place a sample inside it -- and can reach a magnetic field of \qty{2.7}{\tesla} with \qty{3}{\ampere}. When strong magnetic field is required in any direction, a way to reduce the size of the coils is to design a vector magnet where the sample is placed inside one of the bobbin and the other two bobbins are very close or attached to it.\\

Our model is in good agreement with experimental measurements, though in some cases we observe a significant off-set, mainly due to unknown Hall probe sensitivity and imprecise positioning, as well as stray magnetic field sources. As a consequence, for magnetic fields below approximately \qty{10}{\milli\tesla}, our predictive accuracy is limited. These deviations are also attributed to the inhomogeneity of the bobbins radius and height, which could be mitigated by designing thicker side plates. For applications requiring highly stable magnetic fields, an upgrade to persistent-mode operation~\cite{Patel2019} could be considered. Overall, our system offers a flexible and adaptable platform for various experimental setups that demand relatively strong, tunable magnetic field intensity and direction at cryogenic temperatures.

\begin{acknowledgments}
We would like to thank Alexandra Bernasconi and Frederick Hijazi for sharing their expertise on small superconducting coils, Jason Mensingh and Olaf Benningshof for their valuable insights on thermalization, and Raymond Schouten for help with the safety measures. We would also like to acknowledge Brammert Habing, Emanuele Urbinati, and Luca Mastrangelo for experimental support. This work is financially supported by the Netherlands Organization for Scientific Research (NWO) as part of a Vici grant (VI.C.222.024). Y.Y.\ gratefully acknowledges funding from the European Union under a Marie Sk\l{}odowska-Curie fellowship.
\end{acknowledgments}

\section*{Author declarations}

\subsection*{Competing Interests}
The authors declare no competing interests.

\subsection*{Author contributions}
\textbf{Gaia Da Prato}: Planning (lead), Calculations (lead), Design (lead), Assembling (lead), Measurements (lead), Data analysis (lead), Writing - original draft (lead), Writing - review and editing (lead).\\
\textbf{Yong Yu}: Design (supporting), Assembling (supporting), Writing - review and editing (supporting).\\
\textbf{Ronald Bode}: Machining (lead), Design (supporting), Writing - review and editing (supporting).\\
\textbf{Simon Gr\"{o}blacher}: Planning (lead), Supervising (lead), Writing - review and editing (supporting).

\section*{Data Availability Statement}
The scripts and designs are available on \href{https://github.com/GroeblacherLab/VectorMagnet}{GitHub}.

\newpage

\appendix

\section{Finite thickness current loop}
\label{app:calc_magnetic}

We consider a single loop carrying a current $I$, with a finite thickness $t$, an internal radius $r_0$, and an external radius $r_0+t$, as depicted in Figure~\ref{fig:calculation}(a). In the case of a flat 2D loop the magnetic field is then given by
\begin{equation}
\label{eq:finite_thickness}
B(z) = \frac{\mu_0 I}{2 t} \int_{r_0}^{r_0+t}\frac{r^2}{(r^2+z^2)^{3/2}}\mathrm{d}r,
\end{equation}
where we integrate the magnetic field contribution over infinitely thin loops with radii ranging from $r_0$ to $r_0+t$. By introducing the dimensionless parameters $x = t/r_0$ and $\tilde{x} = r/r_0$, we can rewrite Eq.~\eqref{eq:finite_thickness} as
\begin{equation}
B(z) = \frac{\mu_0 I}{2 r_0}\frac{1}{x}\int_{1}^{1+x} \frac{\tilde{x}^2}{(\tilde{x}^2+(z/r_0)^2)^{3/2}} \mathrm{d}\tilde{x}.
\end{equation}
Since for all relevant cases $t \ll r_0$, we can approximate the integral as
\begin{equation}
B(z) =\mathrm{lim}_{x\rightarrow 0}\frac{1}{x}\int_{1}^{1+x} f(\tilde{x}, z)\mathrm{d}\tilde{x} = f(1, z) = \frac{\mu_0 I}{2}\frac{r_0^2}{(r_0^2+z^2)^{3/2}}.
\end{equation}
This is equivalent to the expression for the magnetic field of a single loop (Eq.~\eqref{eq:single_loop}) with $r=r_0$. Using similar arguments, this approximation also holds in the 3D case. 

\section{Protection circuit calculation}
\label{app:protection_circuit}

In the low-current regime, where no current $I$ is flowing through the diodes, the supplied voltage is given by
\begin{equation}
V = (R_\mathrm{L} + R_\mathrm{C} + R_\mathrm{S} + R_\mathrm{B})\cdot I,
\end{equation}
where $R_\mathrm{L}$ is the resistance of the long cable from the protection box to the current source, $R_\mathrm{S}$ is the resistance of the short cable from the top of the DR to the box, $R_\mathrm{C} = \qty{0.5}{\ohm}$ is the resistance in the protection circuit, and $R_\mathrm{B}$ is the resistance of all the wires inside the DR (cf.\ the schematic in Figure~\ref{fig:wiring}(c).) We assume $R_\mathrm{B} = \qty{0}{\ohm}$ when the bobbin is superconducting, although there is a small residual resistance from the copper wires in the upper parts of the DR. The total resistance measured by the power supply is the sum of all these resistances
\begin{equation}
R = R_\mathrm{L} + R_\mathrm{C} + R_\mathrm{S} + R_\mathrm{B}.
\end{equation}
At low temperatures, when no quenching occurs, we typically measure $R = \qty{1.06}{\ohm}$. Given that the short cable is approximately \qty{1}{\meter} long and the long cable is around \qty{6}{\meter} long, we assume $R_\mathrm{S} = R_\mathrm{L}/6$, yielding $R_\mathrm{L} = \qty{0.48}{\ohm}$ and $R_\mathrm{S} = \qty{0.08}{\ohm}$. When a quench occurs, the measured resistance initially jumps to $R = \qty{1.5}{\ohm}$ and then continues to increase. This indicates that the resistance inside the DR has increased to at least $R_\mathrm{B} = \qty{0.44}{\ohm}$.

\begin{figure}[htb!]
\centering
\includegraphics[width = 0.9 \linewidth]{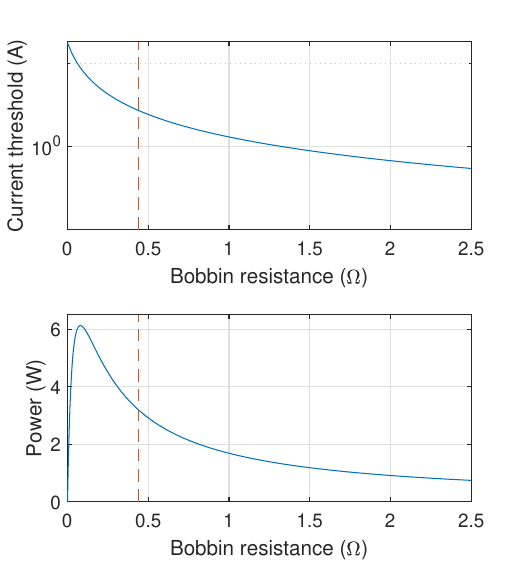}
\caption{Diode current threshold $I_{\mathrm{th}}$ (top plot) and power $P$ flowing in the coil (bottom plot) as a function of the coil resistance $R_\mathrm{B}$. The dashed red line indicates the typical coil resistance as soon as a quench occurs.}
\label{fig:protection_circuit}
\end{figure}

As the current increases, it eventually reaches a threshold value $I_{\mathrm{th}}$, at which point current starts flowing through the diodes
\begin{equation}
I_\mathrm{th} = \frac{V_\mathrm{th}}  {R_\mathrm{S} + R_\mathrm{B}},
\end{equation}
where $V_\mathrm{th}$ is the threshold voltage of two diodes in series, which is around \qty{1.4}{\volt}. The top panel of Figure~\ref{fig:protection_circuit} shows $I_\mathrm{th}$ as a function of $R_\mathrm{B}$. When the coil is superconducting, we find $I_\mathrm{th} = \qty{17.5}{\ampere}$, meaning the coils can be driven up to this current. In a quench scenario, $I_\mathrm{th}$ quickly drops to \qty{2.69}{\ampere}.  Below this value, the protection circuit does not effectively dissipate the current, allowing $R_\mathrm{B}$ to keep increasing until the threshold condition is met. Even when the threshold voltage is reached, some current still flows through the coil, approximately equal to $I_\mathrm{th}$. The power $P$ dissipated in the bobbin when $I > I_\mathrm{th}$ satisfies
\begin{equation}
\label{eq:threshold_power}
\left\{\begin{array}{l}
P = V_\mathrm{B} \cdot I_\mathrm{th} \\
V_\mathrm{B} = V_\mathrm{th} - R_\mathrm{S} \cdot I_\mathrm{th} \\
I_\mathrm{th} =  \frac{V_\mathrm{th}}  {R_\mathrm{S} + R_\mathrm{B}}\\
\end{array}\right.
\end{equation}
where $V_\mathrm{B}$ is the voltage drop inside the DR. Solving for $P$ we get
\begin{equation}
P = V_\mathrm{th}^2 \frac{R_\mathrm{B}}{(R_\mathrm{B}+R_\mathrm{S})^2}.
\end{equation}
The bottom panel of Figure~\ref{fig:protection_circuit} shows the dissipated power as a function of $R_\mathrm{B}$. We observe that even when $R_\mathrm{B} = 5$ $\Omega$, the power remains on the order of \qty{0.5}{\watt}. This suggests that the protection circuit is insufficient to fully prevent DR heating, although it does help in limiting the power in the coils.

\section*{References}
\nocite{*}
\bibliographystyle{aipnum4-1}

\providecommand{\noopsort}[1]{}\providecommand{\singleletter}[1]{#1}%

\end{document}